\documentstyle[prl,twocolumn,aps,epsf]{revtex}
\begin{document}
\draft
\title{Detection of a flow induced magnetic field eigenmode
in the Riga dynamo facility}
\author{Agris Gailitis, Olgerts Lielausis, Sergej Dement'ev,
Ernests Platacis, Arnis Cifersons}
\address{Institute of Physics, Latvian University\\
LV-2169 Salaspils 1, Riga, Latvia}
\author{Gunter Gerbeth, Thomas Gundrum, Frank Stefani}
\address{Forschungszentrum Rossendorf\\
P.O. Box 510119, D-01314 Dresden, Germany}
\author{Michael Christen, Heiko H\"anel, Gotthard Will }
\address{Dresden University of Technology, Dept. Mech. Eng.\\
P.O. Box 01062, Dresden, Germany}

\date{Submitted to Phys. Rev. Lett., December 10, 1999}
\maketitle
\begin{abstract}
In an experiment at the Riga sodium dynamo facility, a slowly
growing magnetic field eigenmode has been detected over a period
of about 15 seconds. For a slightly
decreased propeller rotation rate, additional measurements showed a
slow decay of this mode. The measured results correspond
satisfactory with numerical predictions for the growth rates and
frequencies.
\end{abstract}


\narrowtext

Magnetic fields of  cosmic bodies, such as the Earth, most of the planets,
stars and even galaxies are believed to be generated by the dynamo effect 
in moving electrically conducting fluids. 
Whereas technical dynamos consist of a number of well-separated electrically
conducting parts,
a cosmic dynamo operates, without any ferromagnetism, 
in a nearly homogeneous medium (for an overview see, e.g., \cite{KR1} and 
\cite{MOF1}).

The governing equation for the magnetic field $\bf{B}$ in an electrically
conducting fluid with conductivity $\sigma$ and the velocity
$\bf{v}$ is the so-called induction equation
\begin{eqnarray}
\frac{\partial {{\bf{B}}}}{\partial t}=curl ({\bf{v}} \times {\bf{B}})
+\frac{1}{\mu_0 \sigma} \Delta {\bf{B}}
\end{eqnarray}
which follows from Maxwell equations and Ohms law. The
obvious solution   $\bf{B}=0$ of this equation may become unstable
for some critical value $Rm_c$ of the magnetic Reynolds number
\begin{eqnarray}
Rm=\mu_0 \sigma L v
\end{eqnarray}
if the velocity field fulfills some additional conditions. 
Here $L$ is a typical length scale, and $v$ a
typical velocity scale of the fluid system. $Rm_c$ depends 
strongly on the flow topology and the helicity of the velocity
field. For
self-excitation of a magnetic field it 
has to be at least greater than one. For typical dynamos as the
Earth outer core, $Rm$ is supposed to be of the order of $100$.

 The last decades have seen an enormous progress
of dynamo theory which deals, in its kinematic version, with
the induction equation exclusively
or, in its full version, with the coupled
system of induction equation and Navier-Stokes equation for the
fluid motion. Numerically, this coupled system of equations has been
treated for a number of more or less realistic models of cosmic
bodies (for an impressive simulation, see \cite{GLRO}).

Quite contrary to the success of dynamo theory,
experimental dynamo science is still in its infancy.
This is mainly due to the large dimensions of
the length scale  and/or the velocity scale which are necessary for
dynamo action to occur. Considering the conductivity of sodium
as one of the best liquid conductors
($\sigma \approx 10^{7} \; (\Omega  \mbox{m})^{-1}$
at 100$^{\circ}$C) one gets
$\mu_0 \sigma \approx 10 \; \mbox{s}/{\mbox{m}^2}$. For a very efficient dynamo
with a supposed $Rm_c=10$ this would amount to
a necessary product 
$L v=1 \; \mbox{m}^2/\mbox{s}$ which is very large for a
laboratory facility, even more if one takes into account
the technical problems with  handling sodium.
Historically notable for experimental dynamo science is the
experiment of Lowes and Wilkinson where two ferromagnetic
metallic rods
were rotated in a block at rest \cite{LOWI}.
 A first liquid metal dynamo experiment quite
similar to the present one was undertaken by some of the authors in 1987.
Although this experiment had to be stopped (for reasons of
mechanical stability) before dynamo action occurred  the
extrapolation of the amplification factor of an applied magnetic
field gave indication for the possibility of magnetic field
self-excitation at higher pump rates \cite{GAI1}. Today, there are
several groups working on liquid metal dynamo experiments. For a
summary we refer to the workshop "Laboratory Experiments on Dynamo
Action" held in Riga in summer 1998 \cite{WOR1}.

After years of preparation and careful velocity profile optimization
on water models, first experiments at
the Riga sodium facility were carried out during November 6-11,
1999. The present paper comprises only the most important results
of these experiments, one of them being the observation of a
dynamo eigenmode slowly growing in time at the maximum
rotation rate of the propeller.
A more comprehensive analysis of
all measured data will be published elsewhere.

\begin{figure}
\epsfxsize=8.0cm\epsfbox{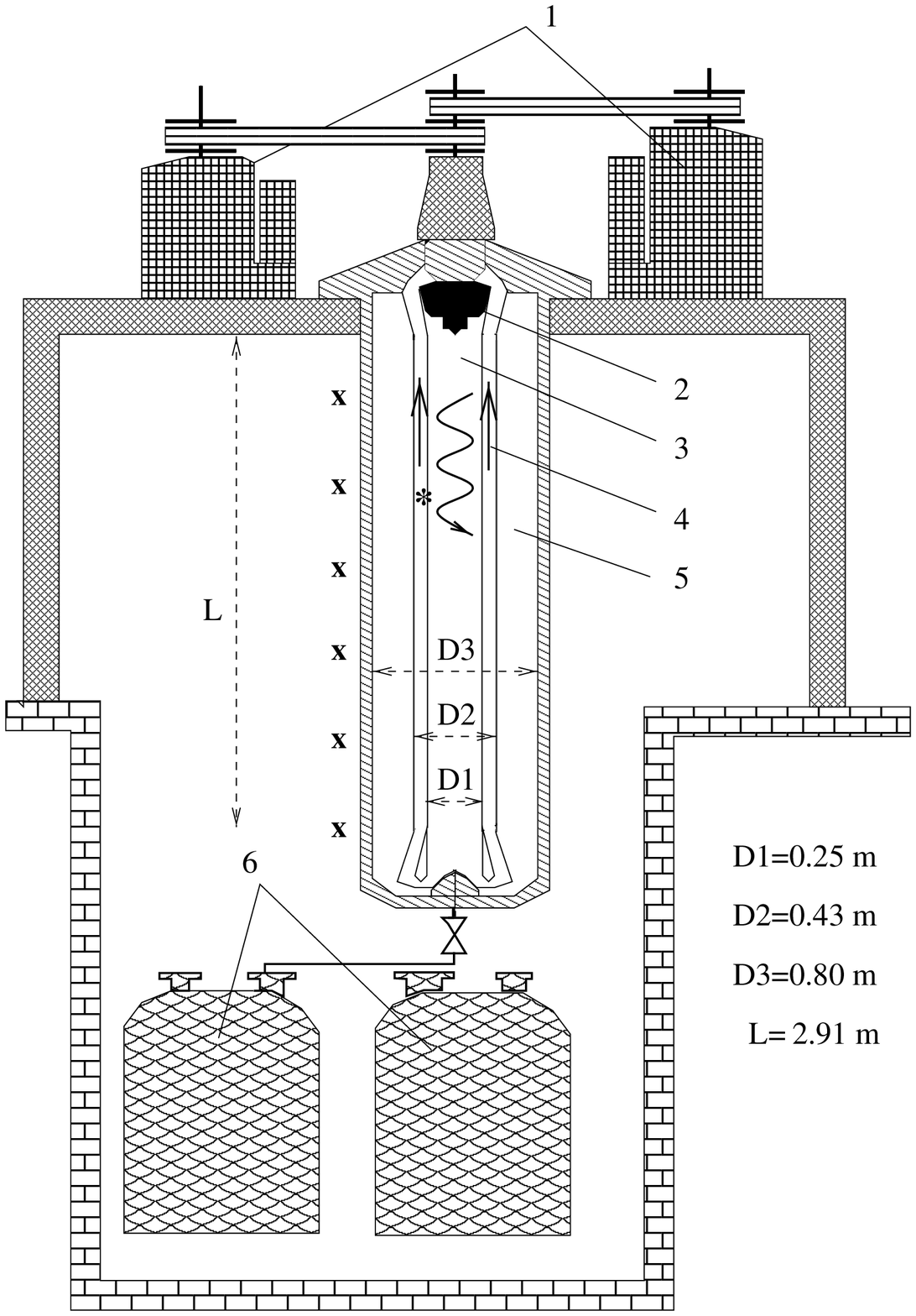}
\vspace{2mm}
\caption{The Riga dynamo facility. Main parts comprising: 
1 - Two motors (55 kW each), 
2 - Propeller, 3 - Helical
flow region, 4 - Back-flow region, 5 - Sodium at rest, 6 - Sodium storage tanks,
$\ast$ - Position of the flux-gate sensor, $\times$ - Positions of the 
six Hall sensors.}
\label{}
\end{figure}

The principal design of the dynamo facility, together with some of
the most important dimensions, is shown in Fig. 1. The main part of
the facility consists of a spiral flow of liquid sodium in an
innermost tube (with a velocity up to 
the order of $15 \; \mbox{m/s}$) with a coaxial
back-flow region and a region with  sodium at rest surrounding it.
The total amount of sodium is $2 \; \mbox{m}^3$. The sodium flow up to $0.6 \; 
\mbox{m}^3/ \mbox{s}$
is produced by a specially designed propeller which is driven 
by two 55 kW motors.

All three sodium volumes play an important role in the magnetic field 
generation
process. The spiral flow within the immobile sodium region amplifies
the magnetic field by stretching field lines \cite{PON}. The back-flow
is responsible  for  a positive  feedback \cite{GAI2}.  The  result  is  an
axially
non-symmetric field (in a symmetric flow geometry!) slowly rotating around
the vertical axis. 
Hence, a low frequency AC magnetic field is expected for this configuration.
Concerning the azimuthal dependence of the magnetic field 
which includes terms of the type
$\exp (i m \varphi)$ with in general arbitrary m it is well-known 
that for those Rm available in this experiment only the
 mode  with $m=1$ can 
play any role \cite{GAI2}.
A lot of details concerning the solution
of the induction equation
for the chosen experimental geometry and
the optimization of the whole facility in general and of the
shape of the velocity profiles in particular can be found in
\cite{GAI2}, \cite{STE1}, and \cite{CHW}.

For the magnetic field measurements we used two different types of
sensors. Inside the dynamo, close to the innermost wall and at a
height of 1/3 of the total length from above, a flux-gate sensor
was positioned. Additionally, 8 Hall sensors were positioned
outside the facility at a distance of 10 cm from the thermal
isolation. Of those, 6 were arranged  parallel to the dynamo axis with
a relative distance of 50 cm, starting with 35 cm from the upper
frame. Two sensors were additionally arranged at different angles.

After heating up the sodium to 300$^{\circ}$C and pumping it slowly through
the facility for 24 hours (to
ensure good electrical contact of sodium with
the stainless-steel walls) various
experiments at 250$^{\circ}$C and around
205$^{\circ}$C at different rotation rates of the propeller
were carried out.

According to our numerical predictions, self-excitation was hardly
to be expected much above a temperature of 200$^{\circ}$C since the
electrical conductivity of sodium decreases significantly with
increasing temperature. Nevertheless, we started experiments at
250$^{\circ}$C in order to get useful information for the later dynamo 
behaviour at lower
temperature, i.e. at higher $Rm$. Although the experiment 
was intended to show
self-excitation of a magnetic field {\it{without any noticeable}}
starting magnetic field, kick-field coils fed by a 3-phase current of
variable low frequency were wound around the module in order to
measure the sub-critical amplification of the applied magnetic
field by the dynamo. This measurement philosophy was quite similar
to that of the 1987 experiment \cite{GAI3} and is based on generation
theory for prolongated flows as the length of our spiral flow exceeds
 its diameter more
then ten times. Generation in such a geometry should
start as exponentially high amplification of some kick-field (known as
convective generation) and should transform, at some higher flowrate, 
into
self-excitation  without any external kick-field.

As a typical example of a lot of measured field amplification curves, Fig.
2 shows the inverse relation of the measured magnetic field to the current 
in the excitation coils for a frequency of 1 Hz and a temperature of 
205$^{\circ}$C versus the rotation rate of the propeller. The two curves 
(squares and crosses) correspond to two different settings of the 3-phase 
current in the kick-field coils with respect to the propeller rotation. An 
increasing amplification of the kick-field can be clearly observed in both 
curves until a rotation rate of about 1500 rpm. These parts of both curves 
point to about 1700 rpm which might be interpreted as the onset of 
convective generation \cite{GAI2}, \cite{STE1}. 
If the excitation frequency would be exactly 
the one the system likes to generate as its eigenmode, the curves should 
further approach the abscissa axis up to the self-excitation point. As 1 Hz 
does not meet exactly the eigenmode frequency the points are repelled from 
the axis for further increasing rotation rates as it is usual for 
externally excited linear systems passing the point of resonance (for this 
interpretation, see also Fig. 5).

\begin{figure}
\epsfxsize=8.6cm\epsfbox{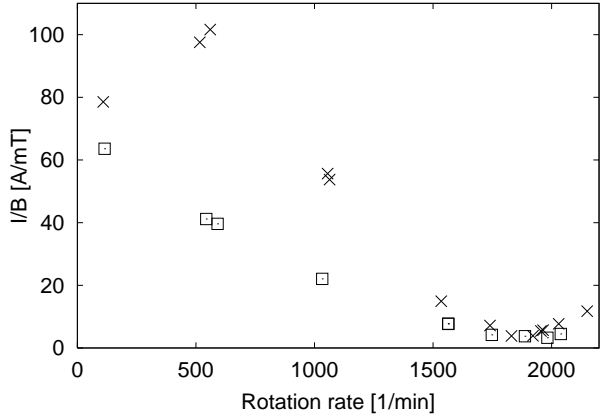}
\caption{Dependence of the magnetic field amplification on the propeller
rotation rate for T=205$^{\circ}$C and f=1 Hz. The ordinate axis shows the
inverse relation of the measured magnetic field to the current in the
kick-field coils. Squares and crosses correspond to two different
settings of the  3-phase 
current in the kick-field coils with respect to the 
propeller rotation.}
\end{figure}

It should be underlined that all points on Fig. 2 except the
rightmost one are
calculated from sinusoidal field records showing the same 1 Hz
frequency as the kick-field. However, the rightmost point at
2150 rpm is exceptional. Let us analyse the
magnetic field signal at this rotation rate in more  detail. Fig 3a
shows the magnetic field measured every 10 ms at the inner sensor
in an interval of 15 s. Evidently, there is a superposition
of two signals.

Numerically, this signal (comprising 1500 data points)
has been analyzed  by means of  a
non-linear least square fit with 8 free parameters according to
\begin{eqnarray*}
B(t)=A_1 e^{p_1 t} \sin{(2 \pi f_1 t +\phi_1)}+
A_2 e^{p_2 t} \sin{(2 \pi f_2 t +\phi_2)}
\end{eqnarray*}
The curve according to this ansatz (which is also shown in Fig. 3a)
 fits extremely well into the data giving the following
parameters (the errors are with respect to a 68.3 per cent
confidence interval):

\begin{eqnarray*}
A_1=&(0.476 \pm 0.004) \; \mbox{mT} ,\; &p_1=(-0.0012 \pm 0.0003) \; \mbox{s}^{-1} \\
f_1=&(0.995 \pm 0.00005) \; \mbox{s}^{-1} , \;&\phi_1=-0.879 \pm 0.012 \\
A_2=&(0.133 \pm 0.001) \; \mbox{mT}  ,\; &p_2=(0.0315  \pm 0.0009) \; \mbox{s}^{-1} \\
f_2=&(1.326 \pm 0.00015)\;  \mbox{s}^{-1}  ,\; &\phi_2=(0.479  \pm 0.009)
\end{eqnarray*}

The positive parameter $p_2=0.0315 \; s^{-1}$ together with
the very small error gives clear evidence for the appearance
of a self-exciting mode at the rotation rate of 2150 rpm.
Fig. 3b shows in a decomposed form 
the two contributing modes, the larger one reflecting
the amplified field of the coils and the smaller one reflecting
the self-excited mode.

\begin{figure}
\noindent(a)\\
\vspace{-10mm}
\epsfxsize=8.6cm\epsfbox{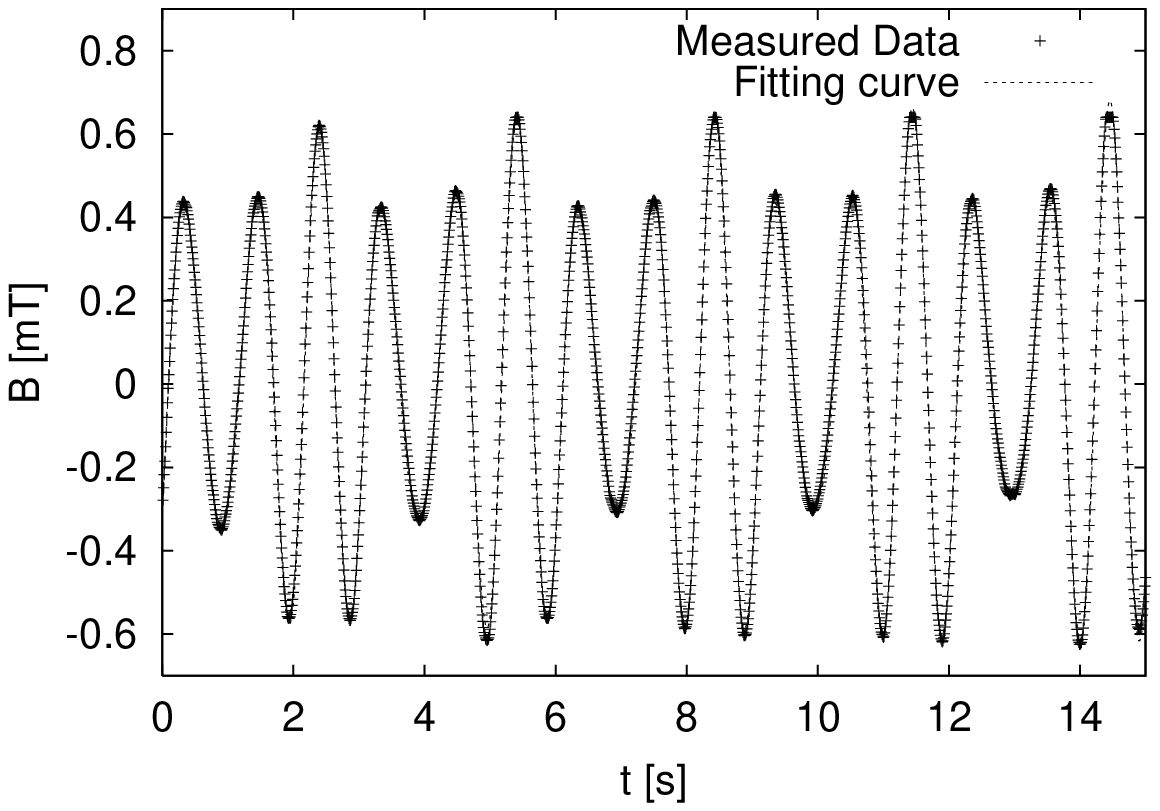}\\
(b)\\
\vspace{-10mm}
\epsfxsize=8.6cm\epsfbox{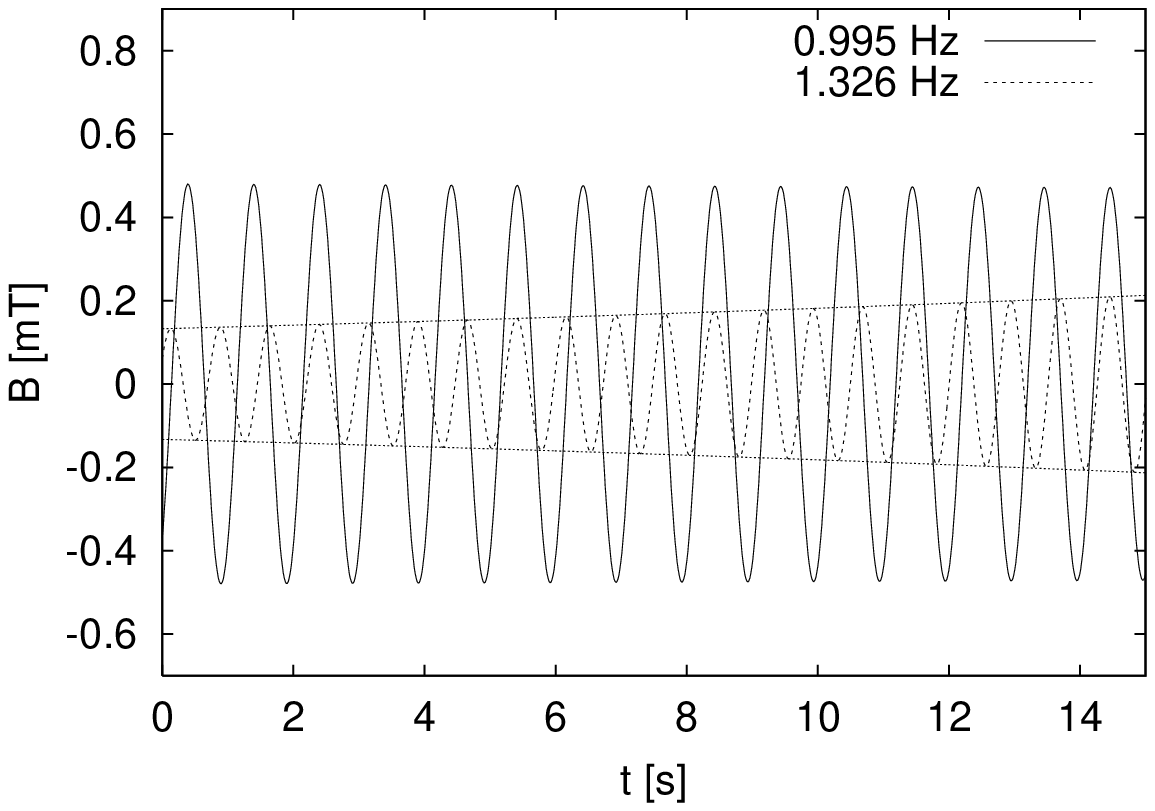}
\caption{Measured magnetic field and fitting curve (a). Decomposition
of the fitting curve into two curves with different frequencies (b).}
\end{figure}

For reasons of some technical problems, this highest rotation rate
could be hold only for 20 seconds  after which it fell
down to 1980 rpm. At that lower rotation rate the coil current was
switched of suddenly. Figure 4 shows the magnetic field behaviour 
at three selected Hall sensors  positioned outside the dynamo.
This mode has a frequency of $f=1.1 \; \mbox{s} ^{-1}$ and a decay 
rate of $p=-0.3 \; \mbox{s} ^{-1}$.
A similar signal was  recorded by the inner fluxgate sensor, too.

It is interesting to compare the frequencies and growth or decay rates
at the two different rotation rates 2150 rpm and 1980 rpm with the numerical
predictions. These are based on the outcomes of a two-dimensional
time dependent code which was described in \cite{STE1}. As input
velocity for the computations an extrapolated velocity field based on
 measurements in water
at two different heights and at three different rotation rates
(1000, 1600, and 2000 rpm) was used. Fig. 5 shows the
predicted growth rates and frequencies for the three
temperatures 150$^{\circ}$C, 200$^{\circ}$C, and 250$^{\circ}$C
 which are different due to
the dependence of the electrical conductivity on temperature. The
two pairs of points in Fig. 5 represent the respective 
measured values.
Having in mind the limitations and approximations of the numerical
prognosis \cite{STE1} the agreement between pre-calculations and
measured values is good, particularly regarding the frequencies of
the magnetic field eigenmode.

\begin{figure}
\epsfxsize=8.6cm\epsfbox{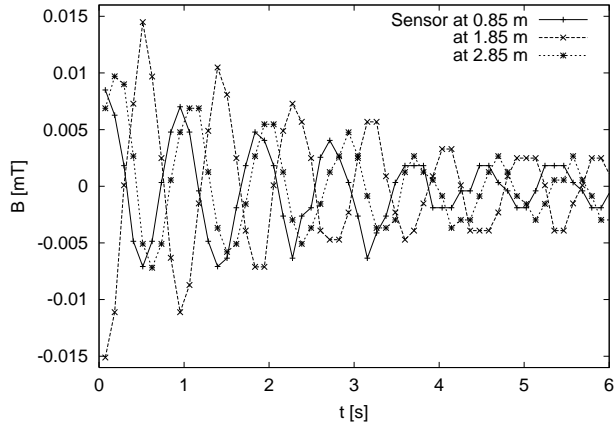}
\caption{Magnetic fields measured at 3 selected  positions
outside the dynamo module after switching off the coil current.}
\end{figure}

The main part of the experiment was originally planned at T=150$^{\circ}$C
where self-excitation with a much higher growth rate was expected.
Unfortunately, the safety rules required to stop the
experiment at T=205$^{\circ}$C since technical problems with the seal of
the propeller axis against the sodium flow-out have been detected. It
is worth to be noted that the overall system worked stable and
without problems over a period of about five days. The sealing
problem needs inspection, but represents no principle problem.

\begin{figure}
\epsfxsize=8.6cm\epsfbox{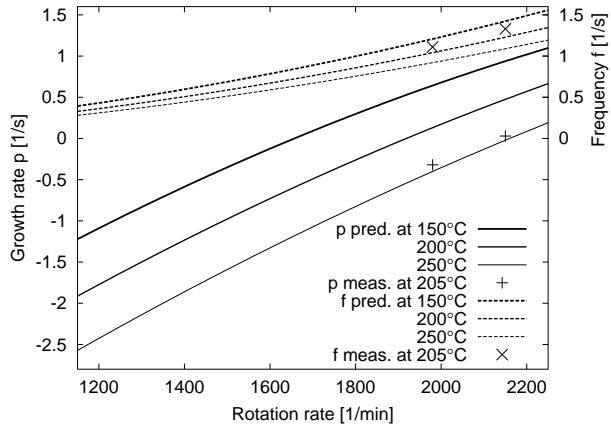}
\caption{Numerical predictions for  growth rates $p$ and
frequencies $f$ of the dynamo eigenmode in dependence on 
the rotation rate for 
three different temperatures, and measured values.} 
\end{figure}

For the first time, magnetic field self-excitation was observed in
a liquid metal dynamo experiment. Expectedly, the observed growth
rate was still very small. The correspondence of the measured
growth rates and frequencies with the numerical prognoses is
convincing. The general concept of the experiment together with
the fine-tuning of the velocity profiles \cite{STE1} have been
proven as feasible and correct. The facility has the potential to
exceed the threshold of magnetic field self-excitation by some 20
per cent with respect to the critical magnetic Reynolds number.
The experiment will be repeated at lower temperature when the
technical problems with the seal will be resolved. For lower
temperature, a higher growth rate will drive the magnetic field to
higher values where the back-reaction of the Lorentz forces on the
velocity should lead to saturation effects.


We thank the Latvian Science Council for support under grant 96.0276,
the Latvian Government and International Science Foundation for support
under joint grant LJD100, the International Science Foundation for support
under grant LFD000 and
Deutsche Forschungsgemeinschaft for support under INK
18/A1-1. We are grateful to W. H\"afele for his interest and
support, and to the whole  experimental  team  for  preparing  and
running the experiment.


%
%
%
%

\end{document}